\documentclass{aa}
\usepackage{graphicx}       
\usepackage{natbib}

\def\H0{{\rm ~km~s^{-1}~Mpc^{-1}}}

\def\ergcms{{~erg~cm$^{-2}$~s$^{-1}$}}

\def\kms{km~s$^{-1}$}
\def\micron{$\mu$m}

\def\nh{{$N_{\rm H}$}}
\def\hi{\ion{H}{I}}

\def\ha{{H$\alpha$}}
\def\hb{{H$\beta$}}

\def\pb{{Pa$\beta$}}
\def\pg{{Pa$\gamma$}}
\def\pd{{Pa$\delta$}}
\def\p9{{Pa$9$}}

\def\ci{[\ion{C}{I}]}
\def\feii{[\ion{Fe}{II}]}

\def\oiii{[\ion{O}{III}]}
\def\pfeii{\ion{Fe}{II}}
\def\phei{\ion{He}{I}}   
\def\pheii{\ion{He}{II}}   
\def\pii{[\ion{P}{II}]}
\def\poi{\ion{O}{I}}   
\def\sii{[\ion{S}{II}]}
\def\siii{[\ion{S}{III}]}
\def\sviii{[\ion{S}{VIII}]}

\def\fullfeii{[\ion{Fe}{II}]$\,\lambda 1.2567\mu$m}

\def\fullpopoi{\ion{O}{I}$\,\lambda$8446}
\def\fullpiroi{\ion{O}{I}$\,\lambda 1.1287\mu$m}

\def\L2-10{L$_{\rm 2-10keV}$}

\def\.25{0.25 keV\thinspace}

\def\d19{D$\,\leq\,$19~Mpc}

\begin{document}

\title{NGC 5506 Unmasked as a Narrow Line Seyfert 1:}
\subtitle{A Direct View of the Broad Line Region using Near-IR 
Spectroscopy{\thanks{Based 
on observations made with the Italian Telescopio Nazionale Galileo (TNG)
operated on the island of La Palma by the Centro Galileo Galilei of the INAF (Istituto
Nazionale di Astrofisica) at the Spanish Observatorio del Roque de los Muchachos of the
Instituto de Astrofisica de Canarias }}
}
\author{N. M. Nagar\inst{1} \and E. Oliva\inst{1,2} \and A. Marconi\inst{1}
        \and R. Maiolino\inst{1}}
\institute{INAF, Arcetri Observatory, Largo E. Fermi 5,
             Florence 50125, Italy 
           \and 
           INAF, Telescopio Nazionale Galileo, Aptdo de Correos, 565, 
           38700 Santa Cruz de La Palma, Canary Islands, Spain 
          }
\offprints{N. Nagar; neil@arcetri.astro.it}

\date{Received May 10, 2002; accepted July 13, 2002}

\abstract{
This letter presents incontrovertible evidence that NGC~5506 is a 
Narrow Line Seyfert 1 (NLSy1). 
Our new 0.9--1.4\micron\ spectrum of its nucleus clearly shows the
permitted \fullpiroi\ line (with full width at half maximum $<\,$2000~\kms) 
and the `1 micron \pfeii\ lines'.
These lines can only originate in the optically-thick broad line region 
(BLR) and, among Seyfert nuclei the latter series of lines are seen 
only in NLSy1s.
The obscuration to the BLR, derived from a rough estimate of the 
\fullpiroi/\fullpopoi\ ratio
and from the reddening of the near-IR Paschen lines, is A$_{\rm V}\,>\,5$. 
Together, these results make NGC~5506 the first identified case of 
an optically-obscured NLSy1. 
This new classification helps explain its radio to X-ray properties, which 
until now were considered highly anomalous. 
However, interesting new concerns are raised: e.g., NGC~5506 is unusual in
hosting both a `type 1' AGN and a nuclear water vapor megamaser.
As the brightest known NLSy1, NGC~5506 is highly suitable for study at 
wavebands less affected by obscuration.

} 

\titlerunning{NGC 5506: A Narrow-Line Seyfert 1}
\authorrunning{Nagar, N. M. et al.}

\maketitle

\section{Introduction}

The Seyfert nucleus of NGC~5506 has resisted a clear type classification 
within Seyfert galaxies, and there is a long standing debate on whether it 
is an intermediate type~1 (broad \ha\ directly visible) or type~2 
(broad \ha\ not directly visible) Seyfert.
The presence of `broad' \pb\ has been reported
by \citet{blaet90}, \citet{rixet90}, and \citet{ruiet94}, but
\citet{gooet94} found that the `narrow' line emission profiles
become broader at longer wavelengths and suggested that the
`broad' \pb\ was the strong, highly reddened wings of this profile.
Based on data available at that time \citet{gooet94} interpreted
the broadening of emission lines with wavelength as due to obscuration
of the inner parts of the narrow line region. \citet{morwar85} reported
a marginal detection of \fullpopoi, characteristic of Seyfert 1s, and 
suggested the presence of high-density optically thick gas.
At odds with other type 2 objects, the nucleus of NGC~5506 is dominated by 
a bright compact core at all near-IR wavelengths and 60\% of the J-band 
(1.25\micron) flux in its central few arcsec is non-stellar in origin 
\citep{oliet99,aloet01}.
In the hard X-ray it is one of the most luminous and brightest Seyferts
in the local universe \citep[\L2-10$\,\sim\,10^{43}$;][]{mus82} 
and its obscuring column 
\citep[\nh$\,= 3.4\,\times\,10^{22}$ cm$^{-2}$;][]{baset99} 
is intermediate between typical values for Seyfert 1s and 2s.
Nuclear water vapor masers, a property highly correlated with 
a type 2 spectral classification \citep{braet96}, 
have been detected towards its nucleus \citep{braet94}. 

The host galaxy causes additional complications. The galaxy disk is close to 
edge on ($i\,=\,70{\degr}$), and dust in the galaxy disk is responsible 
for some or all of the nuclear reddening \citep{veiet97,ima00}.
NGC~5506 is therefore variously treated as a type 1.9 or type 2 Seyfert 
in the literature and in either case is usually an outlier among the 
members of its class.

In this letter we report on near-IR spectroscopy of NGC~5506, which
unequivocally identifies it as a Narrow Line Seyfert~1 (NLSy1). 
In a NLSy1 the broad line region (BLR) is directly visible with the BLR 
emission lines having widths typically $\leq\,2000\,$\kms, significantly 
narrower than those in classical Seyfert 1s. NLSy1s show several anomalous 
properties, most notably in the X-ray 
\citep[for a nice overview of these see][]{veret01} and explanations for
these include accretion rates close to the Eddington rate (implying lower 
black hole masses than other Seyferts) or a view to the AGN along its axis.

\section{Observations and Data Reduction}
Data were obtained on 17 January 2001 using NICS, the near-IR 
imager and spectrograph on the 3.5~meter Telescopio Nazionale Galileo (TNG).
We used the low-resolution IJ grism and a slit width of 
0\farcs75, yielding a dispersion of 5.7\AA/pix (resolution 16\AA)
over the wavelength range 0.89--1.46$\,$\micron, 
and a scale of 0.25{\arcsec}/pixel along the slit. 
The slit was centered on the near-IR peak and rotated
to position angle 140{\degr} North of East.
Total integration time on source was 20~min, consisting of four 5~min
exposures in the standard `ABBA' position raster. The observations 
were immediately followed by observations of the star HIP~69160 
using the same instrument setup.
The sky was not photometric and the seeing about 1{\arcsec}.

Data were reduced using standard tasks within IRAF. The spectrum of 
HIP~69160 (stellar type G5V) was corrected for its intrinsic spectral 
shape and then used to correct the atmospheric absorption lines in the 
spectrum of NGC~5506. Absolute flux calibration was derived by previous
observations of photometric standard stars and should be accurate
to about 30\%. Finally, the spectrum was Doppler corrected using a 
recessional velocity of 1815 \kms\ as derived from \hi\ observations
\citep{vauet91}. The final resolution was $\sim$16~\AA\ over the full 
wavelength range.

\section{Results}

\begin{figure*}
\begin{center}
\resizebox{6in}{!}{
\includegraphics[bb=150 16 596 784,angle=-90]{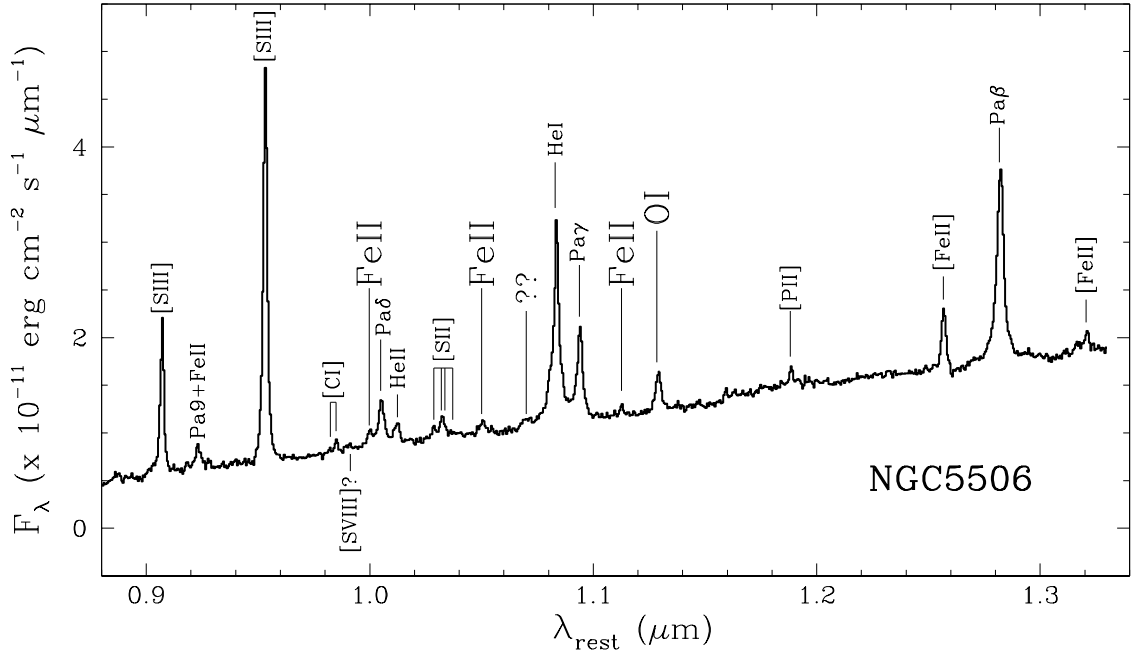}}
\end{center}
\vspace{-0.2in}
\caption{NICS/TNG spectrum of NGC~5506 in an 
0{\farcs}75$\,\times\,$1{\farcs}25 nuclear aperture. Most
detected lines are labelled.
}
\label{figspec}
\end{figure*}

\begin{table}
\label{tabn5506}
\begin{center}
\caption{Line fluxes in NGC~5506}
\begin{tabular}{lrrll}
\hline
Line & Obs. & Eg.             & A$_\lambda$/ & Comments \\
     & flux & Dered.          & A$_{\rm V}$  &   \\
(1)  & (2)  & (3)             & (4)          & (5)  \\
\hline
\poi\  0.8447      &$<$6: &$<$27: & 0.55 & Morris \& Ward  \\
\hline
\siii\  0.9069     &  39  & 105  & 0.48 & \\
\p9$+$\pfeii\ 0.923 
                  &   9  &  23  & 0.47 & blend \\
\siii\  0.9531     & 125  & 280  & 0.44 & includes Pa8 \\
\ci\  0.9850       &   3  &   6  & 0.42 & \\
\sviii\  0.9913    &   1? &   2? & 0.41 & marginal det. \\
\pfeii\  0.9997    &   6  &  12  & 0.41 & \\
\pd\  1.0049       &  18  &  34  & 0.40 & case-B = 34 \\
\pheii\  1.0123    &   8  &  15  & 0.40 & \\
\sii\  1.033       &  12  &  21  & 0.38 & \\
\pfeii\  1.0501    &   6  &  10  & 0.37 & \\            
?? 1.070          &   5  &   8  & 0.36 & seen in NLSy1s \\
\phei\  1.083      &  74  & 110  & 0.35 & \\ 
\pfeii\  1.0863    &   8? &  12  & 0.35 & blend with \phei\ \\
\pg\  1.0938       &  36  &  54  & 0.35 & case-B = 55 \\
\pfeii\  1.1126    &   1: &   1: & 0.35 & marginal det.\\            
\poi\  1.1287      &  14  &  19  & 0.33 & \\
\pii\  1.1882      &   4  &   5  & 0.30 & \\
\feii\  1.2567     &  20  &  21  & 0.27 & \\
\pb\  1.2818       & 100  & 100  & 0.26 & case-B = 100 \\
\feii\  1.3206     &   6  & ... & 0.25 & \\
\hline
\pb\  flux         &  84  & 280  &      & 10$^{-15}$erg/cm$^2$/s \\
\hline
\end{tabular}
\end{center}
{\small{Column~(2) lists the flux measured in our
 0{\farcs}75$\,\times\,$1{\farcs}25 nuclear aperture,
 relative to a \pb\ flux of 100, except for the \poi~0.8446 line 
 which was measured in a 1{\farcs}5 $\times$ 10{\arcsec} 
 nuclear aperture (Morris \& Ward 1985).
 Entries marked with a ``?'' or ``:'' are uncertain.
 The absolute flux of the \pb\ line (accurate to $\pm$30\%) in
 units of 10$^{-15}$\ergcms\ is listed in the last line of the table.  
Column~(3) lists the measured flux after dereddening by an illustrative
 extinction of A$_{\rm V}$=5 (see text), relative to a (dereddened) 
 \pb\ flux of 100.  
Column~(4) lists the A$_\lambda$/A$_{\rm V}$ values, derived from the
 standard extinction curve, used to calculate column~(3).  
The expected relative fluxes for the first three lines in the Paschen 
 series, assuming Case~B recombination, are listed in column~(5).
}}
\end{table}

The final nuclear spectrum 
of NGC~5506 is shown in Fig.~1 with emission-line fluxes listed in Table~1. 
The main result of this paper
is based on our clear detection of the {\textit{permitted}} \fullpiroi\ 
line and the detection of the `1 micron \pfeii\ lines'.
The \fullpiroi\ line, along with \fullpopoi, is produced by Ly$\beta$ pumping 
in a Bowen fluorescence mechanism \citep{gra80}. The latter line was 
tentatively detected by \citet{morwar85}.
Both the above \poi\ lines are produced only by high density optically-thick
gas and are usually seen in Seyfert 1s but never in Seyfert 2s
\citep[e.g.][]{morwar85}. 

The `1 micron \pfeii\ lines' at $\lambda0.9997\mu$m, $\lambda1.0501\mu$m, 
$\lambda1.0863\mu$m, and $\lambda1.1126\mu$m are posited to originate in 
BLR clouds. Theoretically, such \pfeii\ lines and their related optical and
UV counterparts are expected in only type~1 objects and to be strongest in 
NLSy1s \citep[e.g.][]{coljol00}. Observationally,
these lines have been previously detected in only six extragalactic objects, 
all NLSy1s: I~Zwicky~1 \citep{rudet00}, Mrk~478 \citep{rudet01}, 
1H~1934, Ark~564, Mrk~335, and Mrk~1044 \citep{rodet02}. Detailed 
discussions on the origin of the lines can be found in these papers.
Two of the four lines are clearly detected in our spectrum (Fig.~1).
The third, \pfeii~$\lambda1.0863\mu$m, is blended with the very strong 
\phei~$\lambda1.083\mu$m line and the fourth, \pfeii$\,\lambda1.1126\mu$m,
is only marginally detected as it is in a region of atmospheric absorption.
A broad emission feature at 1.07\micron, just blue-ward of 
\phei\ $\lambda1.0630\mu$m, is also present. This feature is also seen in 
all six NLSy1s listed above but has not been identified.

The \pb\ line has a broad pedestal (Fig.~2) and is best fit (after deconvolving
the instrumental resolution) by two Gaussians 
with full width half maximum (FWHM) 500~\kms\ and 1800~\kms; the broader line
contains $\sim$53\% of the flux. 
The similarity between the \poi\ (BLR only) and \pb\ (BLR$+$NLR) line profiles 
(Fig.~2) suggests that this double Gaussian fit does not perfectly separate 
emission from the NLR and BLR, respectively. The \feii\ line profile is only slightly
resolved at our instrument resolution, but nevertheless appears different 
from that of \pb\ (Fig.~2). 
These results are different from those of \citet{veiet97} who found that both 
\pb\ and \feii\ have similar profiles, with their wings well fitted with a Lorenzian
rather than a broad Gaussian component. We were unable to satisfactorily fit our line 
profiles following the functional form of their fit. It is likely that the line 
profiles are affected by significant variability. 
Given the FWHM of the broad components of 
Br$\gamma$ (1550$\pm$100 \kms\ from data taken by R. Maiolino in 1995) 
and Br$\alpha$ \citep[1200$\pm$100 \kms;][]{lutet02}, it appears that the width of the
broad component of the near-IR permitted lines does not increase with wavelength,
suggesting that the \fullpiroi\ and \pb\ profiles trace the bulk of the BLR rather 
than only the outer less obscured part. If this is true, the profile of the \ha\ emission 
from the BLR should be similar to that of \poi\ \citep{gra80}, though the latter line
may be narrower in NLSy1s \citep{rodet02}. 

The observed \pb, \pg, and \pd\ fluxes 
are consistent with those expected when viewing case-B recombination 
through an extinction of A$_{\rm V}\,=$ 5 (cols. 3 to 5 of Table~1). 
That is, for 
the case-B assumption and an error of $\sim$13\% in the \pb/\pd\ ratio,
our data are consistent with extinction A$_{\rm V}\,= 5\,\pm$ 1~mag 
towards the Paschen lines.
The true extinction to the broad line region is expected to be higher
than this illustrative value as our
aperture ($\sim$100~pc at the distance of NGC~5506) includes 
emission from the presumably less-extincted narrow line region. 

A more direct measure of the extinction to the BLR comes from 
the \fullpiroi/\fullpopoi\ ratio, which is reddening sensitive 
(intrinsic value = 1.34). In all the published spectra of NLSy1 the 
observations are compatible with zero relative extinction between the 
two lines \citep{rudet00,rudet01,rodet02}.
The only available data on \fullpopoi\ in NGC~5506 is the spectrum of 
\citet{morwar85,morwar88} where this line is only marginally detected. 
Their spectrum is flux calibrated and partially overlaps with ours. 
From a comparison between both absolute \poi\ fluxes and \poi\ fluxes scaled
to the \siii\ lines{\footnote{A direct comparison of absolute 
fluxes gives \fullpiroi/\fullpopoi\ $\geq\,$2 (higher if we consider that 
our narrower slit captured less light from the seeing-limited BLR). 
A comparison of \poi\ fluxes scaled to the \siii\ lines (after
correcting for differences in the apertures) gives
\fullpiroi/\fullpopoi\ $\sim\,$4 (higher if the aperture corrections 
are not used). 
}} 
we find a lower limit of 2 for the \fullpiroi/\fullpopoi\ flux ratio, 
which translates into a relative reddening A$_{0.8446}\,-\,$A$_{1.1287}\,>$
1. Adopting a standard reddening curve, this gives A$_{\rm V}\,>$ 5.
This result is uncertain due to potential variability, different aperture
sizes, and non-photometric conditions.
Simultaneous observations of both \poi\ lines are therefore highly
desirable.

\begin{figure}
\resizebox{3.6in}{!}{
\includegraphics{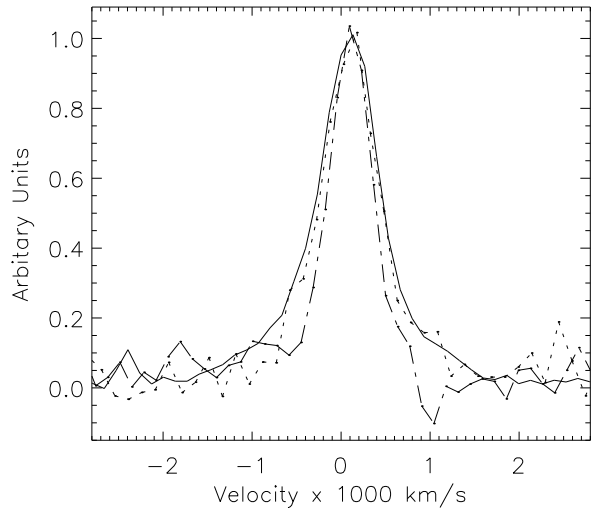}}
\vspace{-0.2cm}
\caption{Comparison of the velocity profiles of \pb\ (solid line), \fullpiroi\ 
 (dotted line), and \fullfeii\ (dash-dot line). The instrumental resolution
 (16\AA\ at all three lines) has not been subtracted out.
}
\label{figprof}
\end{figure}

\section{Discussion}

The currently used classification for NLSy1s  
\citep[from][]{pog00} is:
1.~narrow permitted lines only slightly broader than forbidden lines; 
2.~FWHM(\hb) $<$ 2000~\kms;
3.~\oiii/\hb$\,<\,$3, but exceptions allowed if there is also strong
   [\ion{Fe}{VII}] and [\ion{Fe}{X}] present, unlike what is seen 
   in Seyfert 2s.  \newline
We have shown that the BLR emission is detected in the near-IR and that the 
\poi\ and \pb\ line profiles likely sample the bulk of the BLR.
Thus, with \poi\ (from the BLR only) and \pb\ (from BLR and NLR) line widths 
$<$ 2000~\kms, NGC~5506 directly satisfies the first two conditions. 
The observed \oiii/\hb\ ratio is 7.5 at the nucleus 
and this ratio remains high over most of the extended 
emission-line region \citep{wilet85}. 
If the BLR is highly extincted as our results suggest then the BLR
contribution to the \hb\ flux would likely change the unextincted nuclear 
\oiii/\hb\ ratio to $<\,3$, bringing NGC~5506 into agreement with 
the third condition for classification as a NLSy1. 
A high extinction to the BLR would also explain the lack of strong optical 
\pfeii\ lines as usually seen in NLSy1s.
NGC~5506 shares other properties unique to NLSy1s including the presence 
of the `1 micron \pfeii\ lines' as shown here, a high X-ray luminosity, 
steep X-ray slope, and fast X-ray variability \citep{lamet00}. 
NGC~5506 is now the brightest known NLSy1 and therefore most suited for
studies in wavebands not affected by obscuration.
An important issue raised is whether several other
X-ray bright and highly variable `type 2' Seyferts are, like NGC~5506, 
partially obscured NLSy1s.

Several properties of NGC~5506 still remain, or now become, anomalous.
\citet{matet01} find evidence that NLSy1s have preferentially
lower black hole masses and are accreting at high values of 
L/L$_{\rm Eddington}$.  However, the high central velocity dispersion in 
NGC~5506 \citep[180~\kms;][]{oliet99} though somewhat uncertain,
implies a relatively high black hole mass among Seyferts, if the
scaling between velocity dispersion and black hole mass is valid
among Seyferts \citep{wan02}.
NGC~5506 is also unusual in being a type~1 AGN with a nuclear megamaser.
Both the X-ray column \citep{riset02} and narrow maser lines 
\citep{braet96} are variable, and it may be that the latter are 
produced during periods when the column to nucleus is temporarily higher.

\acknowledgements
We gratefully acknowledge Francesca Ghinassi's help during observations
and the assistance of the NICS and TNG teams.
NN thanks the Centro Galileo Galilei for hospitality during the
writing of this paper.
This work was partially supported by the Italian Ministry for University and
Research (MURST) under grant Cofin00-02-36 and the Italian Space Agency (ASI)
under grant 1/R/27/00.

\clearpage


\begin{thebibliography}{dummy}

\bibitem[Alonso-Herrero et al.(2001)]{aloet01} 
Alonso-Herrero, A., Quillen, A.~C., Simpson, C., Efstathiou, A., 
\& Ward, M.~J.\ 2001, \aj, 121, 1369. 

\bibitem[Bassani et al.(1999)]{baset99} 
Bassani, L., Dadina, M., Maiolino, R., Salvati, M., Risaliti, G., 
della Ceca, R., Matt, G., \& Zamorani, G.\ 1999, \apjs, 121, 473. 

\bibitem[Blanco, Ward, \& Wright(1990)]{blaet90} 
Blanco, P.~R., Ward, M.~J., \& Wright, G.~S.\ 1990, \mnras, 242, 4P. 

\bibitem[Braatz, Wilson, \& Henkel(1994)]{braet94}
Braatz, J.~A., Wilson, A.~S., \& Henkel, C.\ 1994, \apjl, 437, L99. 

\bibitem[Braatz, Wilson, \& Henkel(1996)]{braet96} 
Braatz, J.~A., Wilson, A.~S., \& Henkel, C.\ 1996, \apjs, 106, 51. 



\bibitem[de Vaucouleurs et al.(1991)]{vauet91}
de Vaucouleurs, G., de Vaucouleurs, A., Corwin, H. G., Buta, R. J., 
Paturel, G., \& Fouqu\'{e}, P.  1991, Third Reference Catalogue of Bright 
Galaxies (New York: Springer-Verlag) (RC3)                                                                                                             
\bibitem[Goodrich, Veilleux, \& Hill(1994)]{gooet94} 
Goodrich, R.~W., Veilleux, S., \& Hill, G.~J.\ 1994, \apj, 422, 521. 

\bibitem[Grandi(1980)]{gra80} 
Grandi, S.~A.\ 1980, \apj, 238, 10. 


\bibitem[Imanishi(2000)]{ima00} 
Imanishi, M.\ 2000, \mnras, 313, 165. 

\bibitem[Collin \& Joly(2000)]{coljol00} 
Collin, S.~\& Joly, M.\ 2000, New Astronomy Review, 44, 531 

\bibitem[Lamer, Uttley, \& McHardy(2000)]{lamet00} 
Lamer, G., Uttley, P., \& McHardy, I.~M.\ 2000, \mnras, 319, 949. 

\bibitem[Lutz et al.(2002)]{lutet02} 
Lutz, D., Maiolino, R., Moorwood, A. F. M., Netzer, H., Wagner, S. J., 
Sturm, E., \& Genzel, R. 2002, \aap, submitted

\bibitem[Mathur, Kuraszkiewicz, \& Czerny(2001)]{matet01} 
Mathur, S., Kuraszkiewicz, J., \& Czerny, B.\ 2001, New Astronomy, 6, 321. 

\bibitem[Morris \& Ward(1985)]{morwar85} 
Morris, S.~L.~\& Ward, M.~J.\ 1985, \mnras, 215, 57P. 

\bibitem[Morris \& Ward(1988)]{morwar88} 
Morris, S.~L.~\& Ward, M.~J.\ 1988, \mnras, 230, 639. 


\bibitem[Mushotzky(1982)]{mus82} 
Mushotzky, R.~F.\ 1982, \apj, 256, 92. 



\bibitem[Oliva et al.(1999)]{oliet99} 
Oliva, E., Origlia, L., Maiolino, R., \& Moorwood, A.~F.~M.\ 1999, 
\aap, 350, 9. 



\bibitem[Pogge(2000)]{pog00} 
Pogge, R.~W.\ 2000, New Astronomy Review, 44, 381. 

\bibitem[Risaliti et al.(2002)]{riset02}
Risaliti, G., Elvis, M., \& Nicastro, F. 2002, Publications of
the Astronomical Society of Australia, in press
(astro-ph/0203285)

\bibitem[Rix et al.(1990)]{rixet90} 
Rix, H., Rieke, G., Rieke, M., \& Carleton, N.~P.\ 1990, \apj, 363, 480. 

\bibitem[Rodr{\' i}guez-Ardila et al.(2002)]{rodet02} 
Rodr{\' i}guez-Ardila, A., Viegas, S.~M., 
Pastoriza, M.~G., \& Prato, L.\ 2002, \apj, 565, 140. 

\bibitem[Rudy et al.(2001)]{rudet01} 
Rudy, R.~J., Lynch, D.~K., Mazuk, S., Venturini, C.~C., Puetter, 
R.~C., \& Hamann, F.\ 2001, \pasp, 113, 916. 

\bibitem[Rudy et al.(2000)]{rudet00} 
Rudy, R.~J., Mazuk, S., Puetter, R.~C., \& Hamann, F.\ 2000, 
\apj, 539, 166. 

\bibitem[Ruiz, Rieke, \& Schmidt(1994)]{ruiet94} 
Ruiz, M., Rieke, G.~H., \& Schmidt, G.~D.\ 1994, \apj, 423, 608. 


\bibitem[Veilleux, Goodrich, \& Hill(1997)]{veiet97} 
Veilleux, S., Goodrich, R.~W., \& Hill, G.~J.\ 1997, \apj, 477, 631. 

\bibitem[V{\' e}ron-Cetty, V{\' e}ron, \& Gon{\c c}alves(2001)]{veret01} 
V{\' e}ron-Cetty, M.-P., V{\' e}ron, 
P., \& Gon{\c c}alves, A.~C.\ 2001, \aap, 372, 730. 

\bibitem[Wandel(2002)]{wan02} 
Wandel, A.\ 2002, \apj, 565, 762. 

\bibitem[Wilson, Baldwin, \& Ulvestad(1985)]{wilet85} 
Wilson, A.~S., Baldwin, J.~A., \& Ulvestad, J.~S.\ 1985, \apj, 291, 627. 

\end{thebibliography}
\end{document}